\documentclass[11pt]{article}
\usepackage{graphicx}
\usepackage{amssymb}
\usepackage[super,sort&compress]{natbib}

\input{epsf}
\input{rotate}
\input{psfig.sty}

\bibpunct{(}{)}{,}{s}{,}{,}

\def\blp{\bigg(}
\def\brp{\bigg)}

\def\rv{{\mathbf{r}}}
\def\Rv{{\mathbf{R}}}
\def\B{{\cal{B}}}
\def\H{{\cal{H}}}

\def\aa{\alpha}

\def\NN{{\cal N}}

\def\n{\nu}
\def\a{\alpha}

\begin{document}


{\bf{The shape of a flexible polymer in a cylindrical pore}}\\
\oddsidemargin = 1.0 in
G. Morrison$^{a), b)}$ and D. Thirumalai$^{a), c), d)}$\\
$^{a)}$Biophysics Program, Institute for Physical Science and Technology, University of Maryland, College Park, MD  20742\\
$^{b)}$Department of Physics, University of Maryland, College Park, MD 20742\\
$^{c)}$Department of Chemistry and Biochemistry, University of Maryland, College Park, MD 20742\\
$^{d)}$To whom correspondence should be addressed.  E-mail:  thirum@glue.umd.edu

\oddsidemargin = 0.0 in

\begin{center}
{\bf{Abstract}}
\end{center}
We calculate the mean end-to-end distance ($R$) of a self-avoiding polymer encapsulated in an infinitely long cylinder with radius $D$.  A self-consistent perturbation theory is used to calculate $R$ as a function of $D$ for impenetrable hard walls and soft walls.  In both cases, $R$ obeys the predicted scaling behavior in the limit of large and small $D$.  The crossover from the three dimensional behavior ($D\to\infty$) to the fully stretched one dimensional case ($D\to 0$) is non-monotonic.  The minimum value of $R$ is found at $D\sim 0.46 R_F$, where $R_F$ is the Flory radius of $R$ at $D \to \infty$.  The results for soft walls map onto the hard wall case with a larger cylinder radius.



{\bf{I.  Introduction}}

\qquad Beginning with the observation by {\mbox{Kuhn \cite{Kuhn}}} that polymer coils are {{asymmetric}} even in dilute solutions, a number of studies have characterized the {{anisotropy}} of polymer {\mbox{chains \cite{Lots}}}.  Polymer molecules in good solvents, modeled using the Edwards model, are more {{anisotropic}} than Gaussian polymers because the number of ellipsoidal conformations in self-avoiding chains is far greater than spherical {\mbox{conformations \cite{Eichinger}}}.  The {{anisotropy}} of polymers, which is relevant in a number of applications involving polymer {\mbox{ dynamics \cite{Rheology}}}, becomes even more pronounced in confined spaces.  Nanopores (slits, cylinders and {{gels}}) align the polymer coils and distort their orientations, even when the characteristic confining volume is relatively large compared to the polymer volume \cite{Dave,tenBrinke}.  Confinement-induced alterations in the shape of a polymer is also relevant in biological applications.  For example, a newly synthesized polypeptide chain {{transits}} the ribosome through a roughly cylindrical exit tunnel.  The {{{{extension}}}} is perhaps achieved by an effective stretching {\mbox{force \cite{Klimov}}} $f_s\sim \a \ k_B T/D$, where $T$ is the temperature, $k_B$ is Boltzmann's constant, $D\ (\simeq 1$nm) is the radius of the exit tunnel, and $\a$ is a constant.  The magnitude of $f_s$ that is appropriate to the structure of the tunnel in the ribosome is between 4-10 pN, depending on $\a$, which is large enough to unfold long stretches of proteins at low pulling {\mbox{speeds \cite{Pincus}}}.  Another example is the  encapsulation of a protein in the roughly cylindrical cavity of the {\it{E. Coli}} chaperonin {\mbox{GroEL \cite{GroEL}}}.  In this case, substrate proteins are confined for a duration of time in a nanopore, which can enhance {{unfolding}} rates.  In a very direct application, Tegenfeldt et. {\mbox{al. \cite{PNAS}}} have directly measured the {{{genomic}}} length of DNA molecules by trapping them in cylindrical nanochannels.  

\qquad Motivated in part by the above observations, we consider the behavior of a self-avoiding polymer of contour length $L$, confined to the interior of a cylinder of radius $D$.  We are primarily interested in how the mean end-to-end distance, $R$, of the polymer changes as a function of $D$ and the strength of the interaction between the cylinder and polymer. Daoud and de {\mbox{Gennes \cite{deGennes}}} obtained, using scaling arguments, $R$ when the  interaction with the cylinder is purely repulsive.  As $D\to\infty$, the cylinder has no effect on the mean end-to-end distance $R$, which implies that $R\sim R_F\sim lN^\n$, where $l$ is approximately the size of one monomer, $N$ is the number of monomers, and the Flory exponent $\n=3/(d+2)\simeq 0.6$ in $d=3$ dimensions.  As $D\to 0$, the polymer is effectively confined to $d=1$.  In the confined environment, there are only two relevant length scales, $R_F$ and $D$, so that as {{$D/R_F\to 0$}}, and using the scaling assumption, we can {\mbox{write \cite{deGennes}}}
\begin{eqnarray}
R\sim R_F f(R_F/D).
\end{eqnarray}
As $R_F/D \to \infty$, the chain is stretched in one dimension and becomes rod-like,thus resembling a one-dimensional self-avoiding walk.  The scaling function $f(x)$ takes the form
\begin{eqnarray}
 f(x)\sim\left\{ 
\begin{array}{ll}
 1 & x\to 0 \\ 
 x^m & x\to\infty \\
 \end{array}\right.
\end{eqnarray}
where the unknown exponent $m$ is determined from the condition $R\sim N$ as $x\to\infty$, i.e.  $\n(m+1)=1$, so that $m=2/3$ and {\mbox{thus \cite{deGennes}}}
\begin{eqnarray}
R\sim lN\blp {l \over D}\brp^{{2 \over 3}}\label{prefactor}
\end{eqnarray}
The prefactor in eq. ($\ref{prefactor}$), which is a complicated function of $D$ and the polymer-cavity interactions, is difficult to compute.  In this article, we calculate $R$ for arbitrary values of $D$ by adapting the Edwards-Singh {\mbox{(ES) \cite{ES}}} {{uniform}} expansion method, which has been used in a number of {\mbox{applications \cite{Dave,ES,OtherES}}}.  Note that, without the inclusion of an excluded volume term, the system will converge on a one-dimensional random walk, so that $R\sim l N^{1/ 2}$, with no dependence on $D$.  

\qquad The remainder of the paper is organized as follows:  In section II, we calculate $R$ for a polymer confined to a cylinder with infinite polymer-cylinder repulsion.  These calculations are repeated for soft walls in section III, and the differences between the two systems are determined.  Finally, the effect of short ranged monomer-monomer interactions is briefly described in section IV.

{\bf{II.  Hard Walls}}

\qquad A self-avoiding chain is described by the Edwards {\mbox{Hamiltonian \cite{ES}}} $\beta \H[\rv(s)]={3 / 2l} \int_0^L ds\  \dot {{\mathbf{r}^2}}(s)+\B_2,$ 
where $\B_2=V_0 \int_0^L\int_0^L ds\ ds'\ \delta^{(3)}(\rv(s)-\rv(s'))$ and
 $V_0$ is the strength of the self-avoiding interaction.  Following {\mbox{ES \cite{ES}}}, we replace the true Hamiltonian by a reference Gaussian, $\beta \H_0={3 / 2l_1} \int_0^L ds\  \dot {{\mathbf{r}}}^2(s),$ where the effective step length $l_1$ is determined by the ES method.  We write $\beta \H = \beta \H_0+\B_1+\B_2$\ where $\B_1=3(1/l-1/l_1)/2 \ \int_0^L ds \ \dot \rv^2(s).$  
We {\mbox{find \cite{Dave,ES}}} $R^2 \equiv <\Rv^2>=\int d\rv_0d\rv_L\int{\cal{D}}(\rv(s))\ (\rv_L - \rv_0)^2\exp(-\beta \H[\rv(s)])  \sim \ <\Rv^2>_0-\Delta+O(\B_i^2),$ with
\begin{equation}
\label{selfconsistent}
\Delta=\blp<\Rv^2\ \B_1>_0-<\Rv^2>_0<\B_1>_0\brp +\blp <\Rv^2\ \B_2>_0-<\Rv^2>_0<\B_2>_0\brp,
\end{equation}
and where $<\dots>_0$ denotes an average over the reference Hamiltonian $\beta \H_0$.  The optimal value of $l_1$ is chosen to satisfy $<\Rv^2>\equiv <\Rv^2>_0$, which is possible only if $\Delta \equiv 0$.  The condition $\Delta= 0$ results in a complicated self-consistent equation for $l_1$.  For the unconfined case, ES {\mbox{showed \cite{ES}}} that higher order terms in the $\B_i$'s merely alter the numerical coefficient of $l_1$ without affecting the Flory scaling laws.  Thus, higher order terms will be ignored in this paper.  For the remainder of the paper, all averages are taken with respect to the reference Hamiltonian $\H_0$, so the subscripts on the brackets will be dropped.  We also define $S_i \equiv <\Rv^2\ \B_i>-<\Rv^2><\B_i>$.  

\qquad To calculate $\Delta$, the Green's function for the reference Hamiltonian in a cylinder needs to be determined.  Because of the infinitely repulsive polymer-cylinder interaction the Green's function vanishes at the walls of the cylinder.  The resulting Green's function obeys the Heat Equation in an infinite cylinder, and the {{solution}} that satisfies the appropriate boundary conditions, in terms of the cylindrical coordinates ${\mathbf{r}}=(\rho,\phi,z)$, {\mbox{is \cite{HeatConduction}}}
\begin{equation}
\label{GF}
G(\rv_0,\rv_L; L)={G_z(z_0,z_L) \over \pi D^2} \sum_{m=-\infty}^\infty\sum_{n=0}^{\infty}\cos\ m(\phi_L-\phi_0){J_m(\aa_{mn} \rho_0/D) \over J_{m+1}(\aa_{mn})}{J_m(\aa_{mn} \rho_L/D) \over J_{m+1}(\aa_{mn})}e^{-\aa_{mn}^2 l_1 L/6 D^2},
\end{equation}
where $G_z(z_0,z_L)=\int_{z_0}^{z_L} {\cal D}(z)\exp(-{3 / 2 l_1} \int_0^L ds\ \dot z^2)$, $J_m(x)$ is the $m^{th}$ Bessel function, and $\alpha_{mn}$ is its $n^{th}$ positive root.  {{Using}} eq. ($\ref{GF}$), $<\Rv^2>$ becomes
\begin{equation}
<{\mathbf{R}}^2>={1 \over 3}L l_1+{2D^2 \over \NN}\sum_n \bigg\{{1 \over \aa_{0n}^2}\blp1-{4 \over \aa_{0n}^2}\brp e^{-\aa_{0n}^2 l_1 L /6 D^2}-{1 \over \aa_{1n}^2}e^{-\aa_{1n}^2 l_1 L/6D^2}\bigg\}\equiv {1 \over 3}L l_1+<{\mathbf{R}}_2^2>,
\end{equation}
where $\NN=\sum_n\exp({-\aa_{0n}^2 l_1 L /6 D^2})/\aa_{0n}^2$.  The transverse term of the end-to-end distance, \mbox{$<{\mathbf{R}}_2^2>$} \mbox{$=<x^2>+<y^2>$}, scales as $<{\mathbf{R}}_2^2>\sim D^2$ as $D\to 0$, implying that $<z^2>=l_1 L/3 \sim D^{2/3}$ as $D\to 0$ (see eq. ($\ref{prefactor}$)).

\qquad Taking a derivative of $R^2$ {\mbox{gives \cite{ES}}} $S_1 =l_1^2({1/ l}-{1/ l_1}){d}(R^2)/dl_1$, and we find
\begin{eqnarray}
\label{boneeq}
S_1={1 \over 3}L l_1^2\blp{1\over l}-{1 \over l_1}\brp\blp 1-{1 \over  \NN} \sum_n\bigg\{ \blp 1-{4 \over \aa_{0n}^2}-{<\Rv_2^2> \over 2 D^2}\brp e^{-\aa_{0n}^2 l_1 L /6 D^2}-e^{-\aa_{1n}^2 l_1 L/6D^2}\bigg\}\brp.\label{Termone}
\end{eqnarray}

\qquad The second term in eq. ($\ref{selfconsistent}$) is more complicated, but is simplified by splitting the averages into confined and unconfined terms.  The unconfined $z$ averages are calculated by completing the square in the exponent after  Fourier transforming {\mbox{$\delta(z(s')-z(s''))$ \cite{Dave,ES}}}.  To compute $S_2$ we define
\begin{eqnarray}
I_k(m,\{n_i\})=\int_0^1 dx\ x{J_k(\aa_{kn_1} x) \over J_{k+1}(\aa_{kn_1})}{J_k(\aa_{kn_3} x) \over J_{k+1}(\aa_{kn_3})}{J_{m}^2(\aa_{mn_2} x) \over J_{m+1}^2(\aa_{mn_2})}\ \  \mbox{and}\qquad\qquad \\
E_k(m,\{n_i\};t,t')=I_k(m,\{n_i\})\exp\blp-{l_1 L\over 6D^2}(\aa_{kn_1}t_<+\aa_{mn_2}|t-t'|+\aa_{km_3}(1-t_>))\brp\label{eeq},
\end{eqnarray}
where the `time ordering' variables, $t_<$ and $t_>$, are
\begin{equation}
\label{ttprime}
 t_<=\left\{ 
\begin{array}{ll}
 t & t\le t' \\ 
 t' & t>t' \\
 \end{array}\right.\qquad , \qquad 
  t_>=\left\{ 
\begin{array}{ll}
 t' & t\le t' \\ 
 t & t>t' \\
 \end{array}\right.\qquad
\end{equation}
with $t=s'/L$ and $t'=s''/L$.  In terms of these quantities, we find
\begin{eqnarray}
\label{btwo}
S_2&=&{2 \over \NN}\sqrt{{6 L^3 \over \pi^3 l_1}}\sum_{m,\{n_i\}}\int_0^1\int_0^1dtdt'{V_0\over \sqrt{|t-t'|}}  \bigg\{-{E_1(m,\{n_i\};t,t') \over \aa_{1n_1}\aa_{1n_3}}\\
&\ &\qquad +  {E_0(m,\{n_i\};t,t') \over \aa_{0n_1}\aa_{0n_3}}\blp1-{2 \over \aa_{0n_1}^2}-{2 \over \aa_{0n_3}^2}-{<\Rv_2^2> \over 2D^2}-{l_1 L \over 6 D^2}|t-t'|\brp\bigg\}.\nonumber
\end{eqnarray}
\begin{figure}\label{1}
\centerline {\hbox{
{\epsfxsize = 8.5cm \epsffile{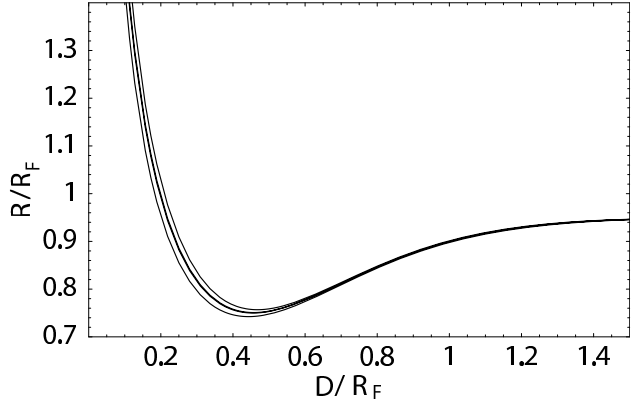}}
}}
\caption{Plots of $R/R_F$ as a function of $D/R_F$, for various values of $L$ and $V_0$.  The lowest curve corresponds to $L/l=1000$ and $V_0/l=0.1$ and the uppermost curve has $L/l=1000$ and $V_0/l=0.5$.  There are two co-incident curves in the middle, one with $L/l=1000$ and $V_0/l=0.1$, the other with $L/l=5000$ and $V_0/l=0.2$.  The minimum for all curves occur near $D_{min}\sim0.46R_F$, with $R_{min}\sim0.75R_F$.} 
\end{figure}
In the limit $D\to\infty$, eq. ($\ref{btwo}$) converges to the Edwards-Singh self consistent equation for the unconfined case \cite{ES}, i.e. 
\begin{equation}
S_1+S_2\sim l_1^2 L \blp{1 \over l}-{1 \over l_1}\brp-2 V_0\sqrt{{6 L^3\over \pi^3 l_1}}=0.\label{noint1}
\end{equation}
Thus, for large $D$, $l_1 \sim (24 V_0^2 l^2 L/\pi^3)^{{1 \over 5}}$ and
$R \sim R_F\sim ({24/ \pi^3})^{{1\over 10}}\ ({V_0 l})^{{1 \over 5}}\ L^{{3 \over 5}}$.
As $D \to 0$, the ground state dominates in eq. ($\ref{GF}$).  Thus, only the $(m,\{n_i\})=(0,0,0,0)\equiv \mathbf{0}$ term in eq. ($\ref{GF}$) makes an appreciable contribution.  In this case, we find that
\begin{equation}
S_1+S_2 \sim {1 \over 3}L l_1^2 \blp{1 \over l}-{1 \over l_1}\brp-{16 V_0 \over 15 D^2} \sqrt{{ l L^5 \over 6 \pi ^3}}I_0({\mathbf{0}})=0\label{noint2}.
\end{equation}
For small cylinder radius, eq. ($\ref{noint2}$) gives $l_1 \sim 0.393 (V_0 l/ D^2)^{{2 \over 3}}L$ and $R\sim 0.627 (V_0 l/ D^2)^{{1 \over 3}} L,$ after $I_0(\mathbf{0})$ is evaluated.  As predicted in eq. ($\ref{prefactor}$), $R$ has the proper scaling form $ND^{-{2 \over 3}}$.  Thus, in both the $D\to0$ and $D\to\infty$ limits, the expected scaling form is recovered, including the predicted dependence of $R$ on $D$.  

\qquad To determine the behavior of $R$ for intermediate values of the cylinder radius, the self-consistent equation ($\Delta=0$) can be numerically solved for $l_1$ as a function of $D$, and $R$ determined using $R=(Ll_1)^{{1 \over 2}}$.  Fig. 1 shows $R/R_F$ for different values of $L$ and $V_0$.  All of the plots are virtually identical, which implies that the only difference between the systems is the numerical value of $R_F$.  As $D/R_F$ exceeds unity, $R\to R_F$ from below (fig. 1).  In accordance with the scaling predictions, $R/R_F$ increases substantially even under moderate squeezing ($D/R_F\sim 0.2$).  Surprisingly, the crossover from the coil state ($D/R_F\gg 1$) to the stretched state ($D/R_F \ll 1$) is non-monotonic.  There is a minimum in $R\sim 0.75 R_F$ at $D_{min}\sim 0.46 R_F$ (Fig. 1).  This behavior, which has been previously observed for polymers confined to a {\mbox{slit \cite{Dave,tenBrinke}}}, is due to confinement-induced anisotropy in the polymer conformations.   Monte Carlo simulations by van Vilet and ten {\mbox{Brinke \cite{tenBrinke}}} show that $R_{min} \sim 0.8 R_F$ in a slit, which shows that confinement in a cylinder squeezes the polymer somewhat more than in a slit.

{\bf{III.  Soft Walls}}

\qquad In real systems, the interaction between the polymer and the cylinder is not infinitely hard.  It is therefore important to calculate $R$ in the case of soft walls.  By soft walls, we mean that the interaction between the polymer and the wall can be represented by a repulsive, non-hard sphere potential, which corresponds to heat conduction with radiative boundary conditions.  In the hard wall case, $G(\rv',\rv_0;L)=G(\rv_L,\rv';L)=0$, with $\rv'=(D,\phi,z)$.  If the walls are soft then the Green's function does not vanish at the boundaries, but {\mbox{satisfies \cite{HeatConductiontwo}}} ${\partial G^S(\rv,\rv';L) / \partial \rho}|_{\rho=D}=C_0 G^S(\rv,\rv';L),$ where $C_0=\infty$ corresponds to the hard wall case.  The Green's function in this case is
\begin{equation}
G^S(\rv_L,\rv_0;L)={G_z \over \pi D^2}\sum_{m=-\infty}^\infty\sum_n {\beta_{mn}^2\over C^2\gamma_{mn}^2} \cos m(\phi_L-\phi_0){J_m(\rho_0 \beta_{mn}/D) \over J_m( \beta_{mn})}{J_m(\rho_L \beta_{mn}/D) \over J_m( \beta_{mn})}\  e^{-\beta_{mn}^2 l_1 L/6D^2},
\end{equation}
where we have defined the dimensionless parameters $C=DC_0$, $\gamma_{mn}=1+(\beta_{mn}^2-m^2)/C^2$, and where the $\beta_{mn}$'s are the positive roots of
\begin{equation}
\label{betaeq}
\beta_{mn} J'_m(\beta_{mn})+CJ_m(\beta_{mn})=0.
\end{equation}
When $C\gg 1$, $\beta_{mn}\sim\aa_{mn}(1-1/C)$, so that $G^{{S}}\sim (1-1/C)G$ by a Taylor expansion.  It can then be shown that, for large C, $R\sim(1+4/5C)R$.  As $D \to \infty$, the results for $R$ for the hard and soft walls coincide.

\qquad For finite $C$, the $\beta_{mn}$'s can not be easily related to the $\aa_{mn}$'s, so we define
\begin{equation}
\NN_S= \sum_{n}{e^{-\beta_{0n}^2 l_1 L / 6D^2} \over \beta_{0n}^2\gamma_{0n}^2}\quad<\bar\Rv_2^2>={2 D^2 \over \NN_S}\sum_{n}{e^{-\beta_{0n}^2 l_1 L/6D^2} \over \beta_{0n}^2\gamma_{0n}^2}\blp 1+{2 \over C}-{4 \over \beta_{0n}^2}\brp-{e^{-\beta_{1n}^2 l_1 L/6D^2} \over \beta_{1n}^2 \gamma_{1n}^2}\blp 1+{1 \over C}\brp^2.\label{newar}
\end{equation}
The $\B_1$ averages are easily computed by taking a derivative of eq. ($\ref{newar}$), as in eq. ($\ref{boneeq}$).  The $\B_2$ averages are tedious to calculate, but give 
\begin{eqnarray}
S_2& =&{2 \over \NN_S}\sqrt{{6 L^3 \over \pi^3 l_1}}\sum_{m,\{n_i\}}{\beta_{mn_2}^2 \over C^4\gamma_{mn_2}^2}\int_0^1\int_0^1 dt dt'{V_0\over \sqrt{|t-t'|}}  \bigg\{-{\bar E_1(m,\{n_i\};t,t') \over \gamma_{1n_1}^2\gamma_{1n_3}^2}\blp 1+{1 \over C}\brp^2 \nonumber\\ 
&\ &\quad + {\bar E_0(m,\{n_i\};t,t') \over \gamma_{0n_1}^2 \gamma_{0n_3}^2}\blp1+{2 \over C}-{2 \over \beta_{0n_1}^2}-{2 \over \beta_{0n_3}^2}-{<\bar\Rv_2^2> \over 2D^2}-{l_1 L \over 6 D^2}|t-t'|\brp\bigg\},\qquad
\end{eqnarray}
where $\bar E_k(m,\{n_i\};t,t')$ is identical to $E_k(m,\{n_i\};t,t')$\ in eq. ($\ref{eeq}$), except that $\alpha_{mn}\to\beta_{mn}$ and $I_k(m,\{n_i\})\to \bar I_k(m,\{n_i\})$, with
\begin{equation}
\bar I_k(m,\{n_i\})=\int_0^1 dx\ x{J_k(\beta_{kn_1} x) \over J_{k}(\beta_{kn_1})}{J_k(\beta_{kn_3} x) \over J_{k}(\beta_{kn_3})}{J_{m}^2(\beta_{mn_2} x) \over J_{m}^2(\beta_{mn_2})}.\qquad\qquad
\end{equation}
As $C\to 0$, it can be shown that eq. ($\ref{betaeq}$) gives $\beta_{00}\sim \sqrt{2C}$.  In the small $D$ limit, for a finite $C_0$, we find $S_1$ converges to the first term in eq. ($\ref{noint2}$), and $S_2 \sim -8 V_0/15 D^2(l_1 L / 6 \pi ^3)^{{1 \over 2}}$,
implying $l_1 \sim 0.230  (l V_0/ D^2)^{{2 \over 3}}L$ and $R \sim 0.526 (  V_0 l / D^2)^{{1 \over 3}}L.$  The scaling laws are unchanged by the softness of the walls; only the numerical coefficients are altered.  
\begin{figure}\label{2}
\centerline {\hbox{
{\epsfxsize = 8.5cm \epsffile{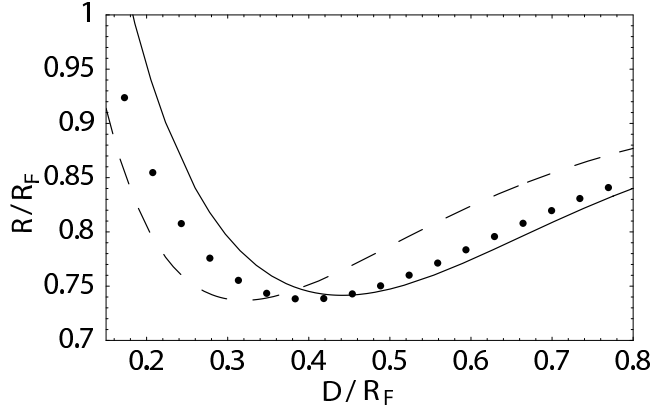}}
}}
\caption{Dependence of $R/R_F$ as a function of $D/R_F$ for various values of $C_0$.  The solid line corresponds to $C_0/l=10^4$, the dotted line is for $C_0/l=0.5$, and the dashed line represents $C_0/l=10^{-5}$.  The values of $L/l$ and $V_0/l$ are 1000 and 0.1, respectively}
\end{figure}

\qquad The numerical solution for $R$ as a function of $D$ with varying $C_0$ shows that $R_{min} \sim 0.75R_F$ as in the hard wall case (fig. 2).   However, as $C_0$ decreases, $D_{min}$ decreases from $\sim 0.46 R_F$ to $\sim 0.32 R_F$ (Fig. 2). Thus, the soft wall behaves as a hard wall with a somewhat larger effective radius, $D_{eff} = D+\delta D$. For $D \ge D_{min}$, a shift of $\delta D \sim D_{min}{(\infty)}-D_{min}(C_0)$ causes $R$ to coincide with the end-to-end distance in the hard wall case.  If we account for this shift, we find that the values of $R$ for both cases  differ at most by $5\%$ for $D\ge D_{min}$.  Because the scaling laws change drastically for small $D$, a simple shift in $D$ is not sufficient to reduce $R$ to the value for hard walls for $D< D_{min}$.

{\bf{IV.  Effect of Monomer-Monomer Interactions on $R$}}

\qquad Typically, there are interactions between monomers besides the universally present excluded volume interactions.  As long as these interactions are short-ranged, the potential between mono\-mers $s_1$ and $s_2$ can be modeled as $\B_3=-\omega\delta(\rv(s_2)-\rv(s_1))$ where $\Delta s=s_2-s_1>0$.  The insertion of this potential into eq. ($\ref{selfconsistent}$) leads to $V_0 \to V_0-\omega \delta(s'-s_1)\delta(s''-s_2)=V_0- \omega \delta(t-t_1)\delta(t'-t_2)/L^2$ in eq. ($\ref{btwo}$), so that for infinitely hard walls, $S_3 \sim \omega (3 / 8 \pi^3 l_1 \Delta s)^{{1 \over 2}}$ as $D\to \infty$, and $S_3 \sim 2\omega I_0({\mathbf{0}})/D^2\ (l_1 \Delta s /6\pi^3)^{{1 \over 2}}$ as $D \to 0$.  Inclusion of monomer-monomer interactions for a polymer confined to soft walls yields the same scaling behavior in both limits, the only change being the numerical coefficients.  Comparison of these scaling laws  with eq. ($\ref{noint1}$) and eq. ($\ref{noint2}$) shows that there is an effective shift in the strength of repulsion due to a monomer attraction, with $\Delta V_0 \sim -\omega \Delta s^{{1 \over 2}} L^{-{5 \over 2}}$ for small $D$, and $\Delta V_0 \sim -\omega \Delta s^{-{1 \over 2}} L^{-{3 \over 2}}$ for large $D$.  In both regimes, the effect of $S_3$ is insignificant compared to $S_2$ for very long chains, provided that $\omega<V_0$.  Short-ranged interactions between all monomers can be computed by assuming that the potential is pairwise additive.  In this case, given a distribution of interactions between the $i^{{th}}$ and $j^{{th}}$ monomers, $\omega_{ij}$, we find that $V_0 \to V_0 -\sum_{i,j} \omega_{ij}\delta^{(3)}(\rv(s_i)-\rv(s_j))$, or $V_0 \to  V_0-\int dtdt'\ V_1(t,t')\delta(\rv(t)-\rv(t'))$, where $V_1(s',s'')=\omega(s',s'')/L^2$.  Because $S_3$ only produces a shift in $V_0$, the addition of this potential will simply reproduce the results in Fig. 1.   for intermediate values of $D$.  Thus, we expect the predicted scaling form to be unaltered for a heteropolymer confined to  a cylinder, provided the effective intra-molecular interaction is not strong enough to induce chain collapse.

{\bf{V.  Conclusions}}

\qquad Inspired by a number of physical situations, we have calculated the dependence of the size of a polymer molecule confined to a cylinder using the Edwards-Singh uniform expansion method.  The theory presented here provides an approximate formula for the dependence of the end-to-end distance $R$ for arbitrary values of the cylinder radius $D$.  The major conclusions of the study are:  ($i$)  The theory yields, in the appropriate limits, the predicted scaling laws for $R$ as a function of $D$.  In particular, the expected scaling function is obtained in the $D\to0$ limit.  The advantage of the theory is that the numerical factors that are difficult to obtain using scaling arguments \cite{ES} have been explicitly calculated.  This  allows for a calculation of $R$ for any value of $D$.  ($ii$) We have calculated $R$ by numerically solving the self-consistent equation.  Surprisingly, we found the crossover from the three dimensional behavior ($D\to\infty$ limit) and the fully stretched limit ($D \to 0$ case) is non-monotonic.  The minimum value of $R\sim 0.75 R_F$ is found at $D\sim 0.46 R_F$ when the wall is infinitely hard.  This is because the wall induces an orienting field that enhances the anisotropy of the polymer.  This effect is greater for an encapsulated polymer in a cylinder compared to slit confinement.  ($iii$)  A direct calculation shows that asymptotic scaling laws are the same for both hard and soft walls.  Any soft wall can be replaced by an equivalent hard wall with a larger cylinder radius, provided the wall-cylinder interaction remains short ranged.

{\bf{Acknowledgments}}\\
\qquad The instructive comments from Prof. Michael E. Fisher are deeply appreciated.  We would like to thank Margaret Cheung and Ed O'Brien for numerous useful discussions.  This work was supported in part by a grant from the National Science Foundation.

\pagebreak

 \end{document}